\documentclass[aps,pra,showpacs,superscriptaddress,numerical,twocolumn,amsmath,amssymb,floatfix,reprint]
{revtex4-1}

\usepackage{graphicx}
\usepackage{dcolumn}
\usepackage{bm}
\usepackage[
	pdffitwindow=true,
	colorlinks=true,
	frenchlinks=false,
        linkcolor=blue,
	anchorcolor=blue,
        citecolor=blue,
        filecolor=blue,
        urlcolor=blue,
        bookmarks=true,
        bookmarksopen=true,
	bookmarksnumbered=true, 
        bookmarksopenlevel=1,
        plainpages=false,
	pdfpagelayout=OnePage,
        pdfpagelabels=true,
	breaklinks
]{hyperref}

\usepackage[per-mode=symbol,separate-uncertainty]{siunitx}
\usepackage[caption=false]{subfig}
\usepackage{braket}
\newcommand{\tens}[2]{%
	\mathbin{\mathop{\bigotimes}\limits_{#1}^{#2}}%
}
\renewcommand\bra[1]{{\langle{#1}|}}
\makeatletter
\renewcommand\ket[1]{%
  \@ifnextchar\bra{\k@t{#1}\!}{\k@t{#1}}%
}
\newcommand\k@t[1]{{|{#1}\rangle}}
\makeatother

\graphicspath{{Figures/}}

\begin{document}

\preprint{APS/123-QED}

\title{Efficient classical simulation of open\\ bosonic quantum systems}

\author{Akseli M\"akinen}
\affiliation{IQM, Keilaranta 19, FI-02150 Espoo, Finland}
\affiliation{QCD Labs, QTF Centre of Excellence, Department of Applied Physics, Aalto University, P.O.~Box 13500, FI-00076 Aalto, Finland}
\affiliation{Research Center for Emerging Computing Technologies (RCECT), National Institute of Advanced Industrial Science and Technology (AIST), 1-1-1 Umezono, Tsukuba, Ibaraki 305-8568, Japan}
\author{Joni Ikonen}
\affiliation{IQM, Keilaranta 19, FI-02150 Espoo, Finland}
\affiliation{QCD Labs, QTF Centre of Excellence, Department of Applied Physics, Aalto University, P.O.~Box 13500, FI-00076 Aalto, Finland}
\author{Takaaki Aoki}
\affiliation{Research Center for Emerging Computing Technologies (RCECT), National Institute of Advanced Industrial Science and Technology (AIST), 1-1-1 Umezono, Tsukuba, Ibaraki 305-8568, Japan}
\affiliation{Department of Physics, The University of Tokyo, 5-1-5 Kashiwanoha, Kashiwa, Chiba 277-8574, Japan}
\author{Jani Tuorila}
\affiliation{IQM, Keilaranta 19, FI-02150 Espoo, Finland}
\author{Yuichiro Matsuzaki}
\affiliation{Research Center for Emerging Computing Technologies (RCECT), National Institute of Advanced Industrial Science and Technology (AIST), 1-1-1 Umezono, Tsukuba, Ibaraki 305-8568, Japan}
\affiliation{NEC-AIST Quantum Technology Cooperative Research Laboratory, National Institute of Advanced Industrial Science and Technology (AIST), Tsukuba, Ibaraki 305-8568, Japan}
\author{Mikko M\"ott\"onen}
\affiliation{QCD Labs, QTF Centre of Excellence, Department of Applied Physics, Aalto University, P.O.~Box 13500, FI-00076 Aalto, Finland}
\affiliation{VTT Technical Research Centre of Finland Ltd., QTF Center of Excellence, P.O. Box 1000, FI-02044 VTT, Finland}

\date{\today}

\begin{abstract}
We propose a computationally efficient method to solve the dynamics of operators of bosonic quantum systems coupled to their environments.
The method maps the operator under interest to a set of complex-valued functions, and its adjoint master equation to a set of partial differential equations for these functions, which are subsequently solved numerically.
In the limit of weak coupling to the environment, the mapping of the operator enables storing the operator efficiently during the simulation, leading to approximately quadratic improvement in the memory consumption compared with the direct approach of solving the adjoint master equation in the number basis, while retaining the computation time comparable.
Moreover, the method enables efficient parallelization which allows to optimize for the actual computational time to reach an approximately quadratic speed up, while retaining the memory consumption comparable to the direct approach.
We foresee the method to prove useful, e.g., for the verification of the operation of superconducting quantum processors.
\end{abstract}

\maketitle

\emph{Introduction.---}In recent years, the potential to revolutionize a spectrum of industries~\cite{Bova_2021}, such as finance~\cite{Egger_2020}, pharmacy~\cite{Cao_2018}, and cybersecurity~\cite{Mosca_2018}, has given a major impetus towards realizations of practical-use quantum computers \cite{Knill_2001, Ladd_2010, Fowler_2012, Nigg_2014, Billangeon2015, Debnath_2016, Ikonen_2017, Rasanen_2021}. The role of classical simulations of such devices in the progress is of great importance. First, accurate classical simulations contribute to the optimization of the control of a quantum processor to maximize the fidelity of the target quantum algorithm. Second, while there has already been claims of achieving the quantum supremacy~\cite{Arute_2019, Zhong_2020}, these experiments need to be verified against fully controllable classical simulations. Third, a fair proof of quantum supremacy requires a comparable effort put on the development of classical simulations of the given experimental setup, which may raise the bar for quantum computers.

Currently, some of the most prominent architectures for programmable quantum computers are based on superconducting technology~\cite{Arute_2019, Mooney_2021}. Such devices, among a range of other quantum technological applications~\cite{OBrien_2009, Pirkkalainen_2013, Zhou_2013}, such as those employing quantum optics~\cite{Zhong_2020, Walmsley_2015} and circuit quantum electrodynamics \cite{Blais_2004, Tan_2017}, can predominantly be considered as bosonic systems. Recent progress in the development of quantum technology, especially in superconducting circuits \cite{Megrant_2012, Barends_2014, Goetz_2016b}, has enabled reaching the limit of very weak interaction between the system under interest and its environment~\cite{Jurcevic_2020, Mooney_2021}. Therefore, the weak interaction limit has become increasingly relevant also for classical simulations.

Typically, the experiments with quantum processors terminate with the measurement of the qubits in their computational basis.
However, a variety of problems of experimental interest use instead measurements of certain other observables, including certain non-commuting observables~\cite{Hacohen_2016}, correlation functions~\cite{Pedernales_2014}, and  dynamical linear response functions~\cite{Rall_2020}.
The Heisenberg picture is the natural picture for the implementation of classical simulations of such schemes. 
The direct approach to solve the dynamics of a given operator in the Heisenberg picture is to transform its adjoint master equation into a set of coupled scalar differential equations corresponding to the number basis, or to some other basis~\cite{Bender_1989, Mista_2001, Razavy_2011}, and to integrate this set numerically.
In this Letter, we refer to the former of these methods to as the conventional approach. Alternative techniques include the so-called coherent state representation~\cite{Dalvit_2006} and the phase space formulation~\cite{Agarwal_1970a, Agarwal_1970b}.
However, in general these approaches are memory-consuming which poses stringent requirements for the utilized computational hardware.

In this Letter, we push the boundaries of classical simulations of weakly dissipative bosonic quantum systems in the Heisenberg picture. The proposed approach is based on expanding the operator under interest in a problem-specific operator basis, mapping the adjoint master equation of the operator into a complex-valued partial differential equation for the generating function of this expansion, performing a simple and well-defined transformation to this dynamic equation, and solving the transformed dynamic equation numerically. In the process, the operator is mapped to a set of complex-valued functions in a fashion that enables storing it memory-efficiently during the simulation. The approach poses no assumptions on the structure or on the initial conditions of the system.

To demonstrate the power of the proposed method, we consider a network of coupled driven dissipative anharmonic oscillators, routinely used as a model of superconducting quantum processors based, for example, on transmon qubits~\cite{Yan_2018, Zhao_2020}. To this end, we first use a well-known property of Markovian master equations to show that the most commonly used master equation for the anharmonic oscillator breaks down already for arbitrarily weak anharmonicity, and consequently derive the proper master equation starting from the first principles. Subsequently, we use the conventional approach and our approach to numerically solve the proper adjoint master equation of the system and compare the memory consumptions and computation times of the methods.

\begin{figure}
  \centering
  \includegraphics[width = \columnwidth]{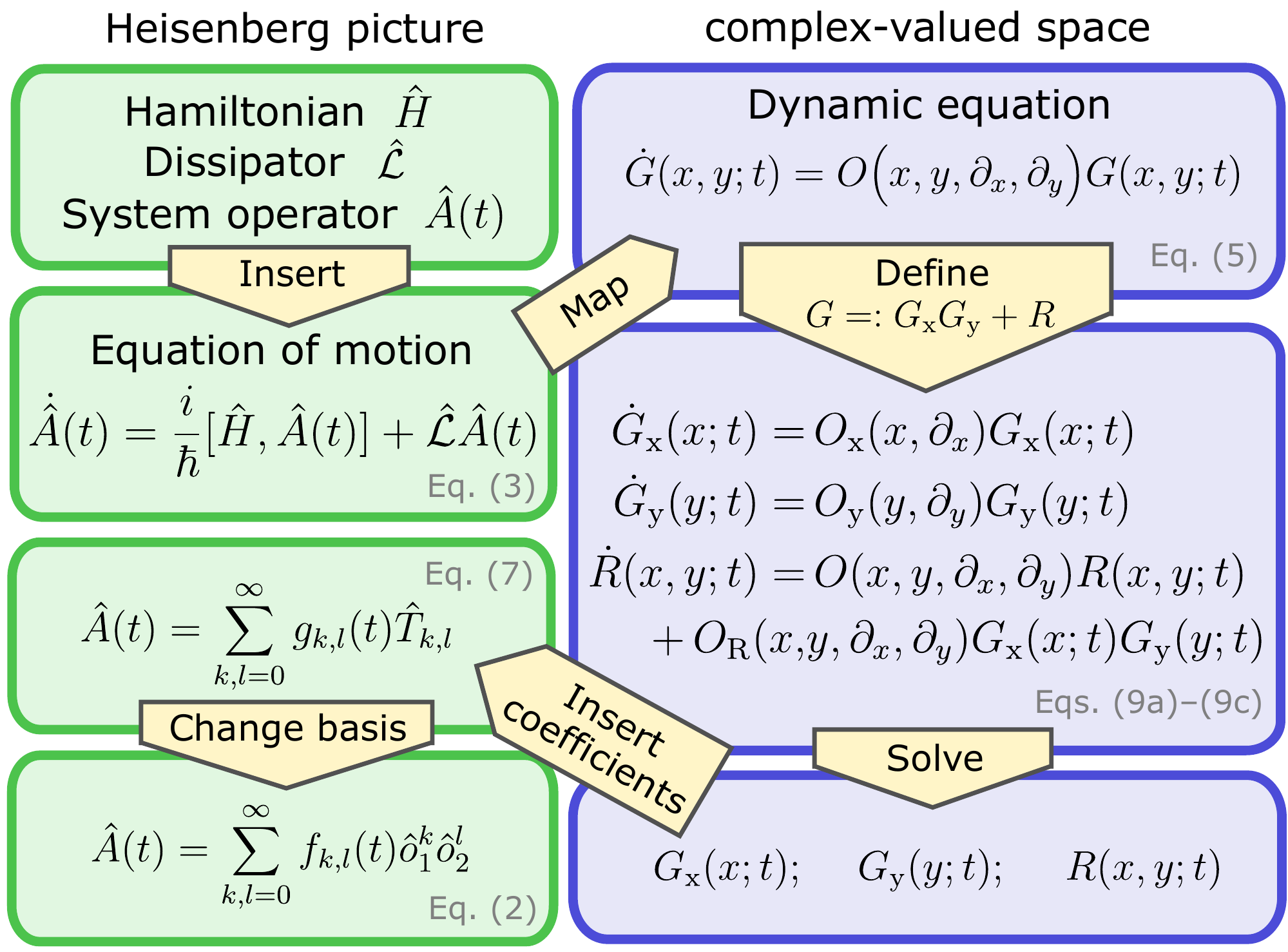} 
  \caption{\label{fig:1} Schematic process chart of the proposed method to solve the dynamics of operators of open bosonic quantum systems. The symbols are defined in the main text in the vicinity of the related equations. For clarity, we have used the short-hand notation $O$ for the temporal evolution operator of the generating function $G$ and similar notations for those of $G_{\text{x}}$, $G_{\text{y}}$ and $R$, and suppressed the temporal dependencies of the Hamiltonian and the temporal evolution operators.}
\end{figure}

\emph{The method.---}In this section, we provide a general description of the proposed method, see Fig.~\ref{fig:1}. For the sake of clarity, we limit the consideration to single-mode systems. However, the method straightforwardly generalizes to an arbitrary number of discrete bosonic modes as we show in~\cite{supp}. Moreover, here we do not construct the dynamics for the general dissipators. Such a treatment is given subsequently when we consider the example system of anharmonic oscillators.

We begin by introducing two elementary operators of the system under interest, $\hat{o}_1$ and $\hat{o}_2$, such that (i)~they obey the commutation relation $[ \hat{o}_2, \hat{o}_1 ]=r\hat{I}$, where $r \in \mathbb{C} \setminus \{0 \}$ and $\hat{I}$ is the identity operator, and (ii)~their monomials, $\{ \hat{o}_1^k \hat{o}_2^l \}_{k,l=0}^{\infty}$, form a complete operator basis. Consequently, we expand the Hamiltonian of the system in the Heisenberg picture in this basis as
\begin{equation}
\hat{H}(t) / \hbar = \sum_{u,v=0}^{\infty} h_{u,v}(t) \hat{o}_1^u \hat{o}_2^v, \label{eq:01}
\end{equation}
and an arbitrary system operator as
\begin{equation}
\hat{A}(t) = \sum_{k,l=0}^{\infty} f_{k,l}(t) \hat{o}_1^k \hat{o}_2^l. \label{eq:02}
\end{equation}
Examples of elementary operators that satisfy the above conditions~(i) and~(ii) are the creation and annihilation operators of a bosonic mode, $\hat{o}_1 = \hat{a}^{\dagger}$ and $\hat{o}_2 = \hat{a}$, for which $r=1$, and the corresponding quadratures, $\hat{o}_1 = \hat{p}$ and $\hat{o}_2 = \hat{q}$, for which $r=i$. The utility in the introduction of these generic elementary operators is clarified below.

Let us consider a generic adjoint master equation of the system~\cite{Breuer_2002},
\begin{align}
\dot{\hat{A}}(t) =\,& \frac{i}{\hbar}[ \hat{H}(t), \hat{A}(t)] \nonumber \\
&+ \sum_n \frac{\kappa_n}{2} \big[ 2 \hat{L}_n^{\dagger} \hat{A}(t) \hat{L}_n - \big\{ \hat{L}_n^{\dagger}\hat{L}_n, \hat{A}(t)\big\} \big], \label{eq:03}
\end{align}
where the dot denotes the temporal derivative, $\hat{L}_n$ is the Lindblad jump operator of the noise channel~$n$, and $\{\hat{O}_1,\hat{O}_2\}=\hat{O}_1\hat{O}_2+\hat{O}_2\hat{O}_1$ is the anticommutator. Operator differential equations are in general difficult to solve directly. To trasform Eq.~(\ref{eq:03}) into a $c$-number differential equation, we insert the expansions of Eqs.~(\ref{eq:01}) and~(\ref{eq:02}) into Eq.~(\ref{eq:03}) and define a complex-valued function $F(x,y;t) = \sum_{k,l=0}^{\infty} f_{k,l}(t) x^k y^l$ that generates the expansion coefficients of operator $\hat{A}(t)$. A lengthy calculation yields the dynamic equation for the generating function~\cite{supp},
\begin{align}
\dot{F}(&x,y;t) \nonumber \\
=\,& i\big[ :H( x, r\partial_x + y; t \big): - :H\big( x + r\partial_y, y; t ): \big] F(x,y;t) \nonumber \\
&+ \sum_n \kappa_n O_n ( x, \partial_x, y, \partial_y, r ) F(x,y;t), \label{eq:04}
\end{align}
where $:*:$ denotes ordering of the variables to the left with respect to the differentiation operators, $H(x,y;t) = \sum_{u,v=0}^{\infty} h_{u,v}(t) x^u y^v$, and $O_n$ is a multivariate function of the coordinates and the differential operators associated with the noise channel $n$. For brevity, we omit the general form of $O_n$. However, in Eq.~(S21) in~\cite{supp} we implicitly provide it for the anharmonic oscillator, and the technique used therein to derive it is also applicable for other systems of interest. Note that the choice of the function basis in the definition of $F$ is arbitrary and different choices lead to different forms of the dynamic equation.

To the best of our knowledge, the adjoint master equation has not previously been mapped to the form of Eq.~(\ref{eq:04}). Note that Eq.~(\ref{eq:04}) can be readily solved numerically. However, we take here a more involved approach by defining a new generating function as $G(x,y;t) = e^{r^{-1}xy} F(x,y;t)$. The idea is to transform the dynamic equation into an almost separable form to enable storing the generating function in terms of univariate functions and a residual function. When discretizing these functions for simulation, the univariate functions and the residual function can be stored more compactly than the untransformed bivariate generating function. The transformed dynamic equation reads
\begin{align}
\dot{G}(x,y;t)=\,& i\big[ :H( x, r\partial_x; t ): - :H( r\partial_y, y; t ): \big] G(x,y;t) \nonumber \\
&+ \sum_n \kappa_n O'_n( x, \partial_x, y, \partial_y, r ) G(x,y;t), \label{eq:05}
\end{align}
where $O'_n = e^{r^{-1}xy}O_n e^{-r^{-1}xy}$. The new generating function $G(x,y;t) = \sum_{k,l=0}^{\infty} g_{k,l}(t) x^k y^l$ can be expressed in terms of the original coefficients as
\begin{equation}
g_{k,l}(t) = \sum_{j=0}^{\min(k,l)} \frac{r^{-j}}{j!} f_{k-j,l-j}(t). \label{eq:06}
\end{equation}
The new generating function corresponds to the operator expansion
\begin{equation}
\hat{A}(t) = \sum_{k,l=0}^{\infty} g_{k,l}(t) \hat{T}_{k,l}, \label{eq:07}
\end{equation}
where the operator basis elements are given by
\begin{equation}
\hat{T}_{k,l} = \sum_{j=0}^{\infty} \frac{(-r)^{-j}}{j!} \hat{o}_1^{k+j} \hat{o}_2^{l+j}, \label{eq:08}
\end{equation}
which can be verified by inserting Eqs.~(\ref{eq:06}) and~(\ref{eq:08}) into Eq.~(\ref{eq:07}).

Next, we introduce three new complex-valued functions according to $G(x,y;t) = G_{\text{x}}(x;t) G_{\text{y}}(y;t) + R(x,y;t)$, which are governed by
\begin{subequations}
\begin{align}
\dot{G}_{\text{x}}(x;t) =\, &\big[ i:H ( x,r\partial_x;t \big): + C ] G_{\text{x}}(x;t), \label{eq:09a} \\
\dot{G}_{\text{y}}(y;t) =\, &-\big[ i:H ( r\partial_y,y;t \big): + C ] G_{\text{y}}(y;t), \label{eq:09b} \\
\dot{R}(x,y;t) =\,&i\big[ :H ( x,r\partial_x;t ): - :H ( r\partial_y,y;t ): \big]R(x,y;t) \nonumber \\
& + \sum_n \kappa_n O'_n ( x, \partial_x, y, \partial_y, r ) G(x,y;t), \label{eq:09c}
\end{align}
\end{subequations}
where $C$ is the separation constant. Suppose that the new generating function is initially separable, that is, $G(x,y;0) = G_{\text{x}}(x;0) G_{\text{y}}(y;0)$ and $R(x,y;0)=0$. From Eq.~(\ref{eq:09c}) it is evident that if $\kappa_n t$ is small for all $n$, then the function $R$ can be approximated as a zero function. Moreover, in the following we find that in the case of a network of damped anharmonic oscillators and larger $\kappa_n t$, $R$ can be approximated as a single-variable function. This implies that the transformation enables storing the operator memory-efficiently as a set of three single-variable functions intead of one two-variable function.

Note that any $G(x,y;0)$ can be expressed as a linear combination of separable functions, e.g.~as
\begin{align}
G(x,y;0) &= \sum_{p,q=0}^{\infty} g_{p,q}(0) x^p y^q \nonumber \\
&=: \sum_{p,q=0}^{\infty} g_{p,q}(0) G^{(p)}_{\text{x}}(x;0) G^{(q)}_{\text{y}}(y;0) \nonumber \\
&=: \sum_{p,q=0}^{\infty} g_{p,q}(0) G^{(p,q)}(x,y;0). \label{eq:10}
\end{align}
Thus, by solving the linear dynamics of each of such initially separable function $G^{(p,q)}(x,y;0)$, we can reconstruct the dynamics of $G(x,y;t)$. The initial expansion coefficients read
\begin{equation}
f_{k,l}(t) = \sum_{j=0}^{\min (k,l)} \frac{(-r)^{-j}}{j!} \sum_{p,q=0}^{\infty} g_{p,q}(0) g_{k-j,l-j}^{(p,q)}(t), \label{eq:11}
\end{equation}
where $g_{k,l}^{(p,q)}(t)$ are the expansion coefficients of $G^{(p,q)}(x,y;t)$ in the chosen function basis $\{ x^k y^l \}_{k,l=0}^{\infty}$. Note that this approach requires solving the dynamics of $N_{\text{thr}}^2$ generating functions, where $N_{\text{thr}}$ denotes the truncation of the Taylor expansion in Eq.~(\ref{eq:10}). This number is reduced to $N_{\text{thr}}$ if the initial operator is sub-, super-, or diagonal in the chosen operator basis, that is,
\begin{align}
\hat{A}(0) &= \sum_{k=\max (0,-d)}^{\infty} f_{k,k+d}(0) \hat{o}_1^k \hat{o}_2^{k+d}, \label{eq:12}
\end{align}
where $d\in \mathbb{Z}$ defines whether the initial operator is sub- ($d<0$), super- ($d>0$), or diagonal ($d=0$). In this case, the initial expansion coefficients read
\begin{align}
f_{k,l}(t) &= \sum_{j=0}^{\min (k,l)} \frac{(-r)^{-j}}{j!} \sum_{p=\max (0,-d)}^{\infty} g_{p,p+d}(0) g_{k-j,l-j}^{(p,p+d)}(t). \label{eq:13}
\end{align}
Examples of operators that possess this favorable property include the number state projectors, the powers and cumulants of $\hat{o}_1 \hat{o}_2$, the operator defining the leakage error from a subspace spanned by number states, and the squeezing operator. Note that the dynamics of each $G^{(p,q)}(x,y;t)$ is independent. Thus, the numerical integration of these transformed generating functions can be parallelized in an ideal fashion.

In summary, the method presented in this section can be used to solve the dynamics of an operator of an open bosonic system as follows: (i)~Determine the initial transformed expansion coefficients $g_{p,q}(0)$ using Eqs.~(\ref{eq:02}) and~(\ref{eq:06}). (ii)~Numerically solve Eqs.~(\ref{eq:09a})--(\ref{eq:09c}) with the initial condition $G^{(p,q)}(x,y;0)=x^p y^q$ for each non-negligible $g_{p,q}(0)$. (iii)~If desired, reconstruct the original expansion coefficients of the operator using Eq.~(\ref{eq:11}).

\emph{Master equation for the anharmonic oscillator.---}In this section, we present a master equation for the anharmonic oscillator, which we subsequently solve to demonstrate the power of our method. To justify solving it instead of the conventionally used master equation, we first argue that the latter is not accurate even for low values of anharmonicity.

It has been shown that the von Neumann entropy production rate $\Pi = \dot{S} - T_{\text{E}}^{-1} \dot{E}$ of any Markovian quantum system is non-negative~\cite{Breuer_2002}, where $S=-k_{\text{B}}\text{Tr}(\hat{\rho}\ln \hat{\rho})$ is the von Neumann entropy of the system, $T_{\text{E}}$ is the temperature of its environment, $E=\langle \hat{H} \rangle$ is the mean energy of the system, $k_{\text{B}}$ is the Boltzmann constant, and $\hat{\rho}$ is the density operator of the system. Consider the commonly used master equation of the anharmonic oscillator~\cite{Drummond_1980, Milburn_1986, Chaturvedi_1991, He_2015, Rossi_2016}
\begin{align}
\dot{\hat{\rho}}(t) =& - i \omega[\hat{a}^{\dagger}\hat{a},\hat{\rho}(t)] - i \eta[(\hat{a}^{\dagger})^2\hat{a}^2,\hat{\rho}(t)] \nonumber \\
&+ \frac{\kappa}{2}(\bar{n}+1)\big[2\hat{a}\hat{\rho}(t)\hat{a}^{\dagger} - \hat{a}^{\dagger}\hat{a}\hat{\rho}(t) - \hat{\rho}(t)\hat{a}^{\dagger}\hat{a}\big] \nonumber \\
&+ \frac{\kappa}{2}\bar{n}\big[2\hat{a}^{\dagger}\hat{\rho}(t)\hat{a} - \hat{a}\hat{a}^{\dagger}\hat{\rho}(t) - \hat{\rho}(t)\hat{a}\hat{a}^{\dagger}\big], \label{eq:201}
\end{align}
where $\omega$ is the angular frequency difference between the two least energetic states, $\eta$ is the anharmonicity, and $\bar{n}=[e^{\hbar \omega/(k_{\text{B}}T_{\text{E}})} - 1]^{-1}$ is the mean number of excitations in the environment in the mode with the angular frequency $\omega$. In~\cite{supp} we show that given a diagonal initial state of the form $\hat{\rho}(0)=\mathcal{N} \sum_{n=0}^{\infty} e^{-cn}|n\rangle \langle n|$, where $\mathcal{N}$ is a normalization constant, for all non-zero values of $\eta$ there exists $c>0$ such that the conventionally used master equation leads to a negative entropy production rate. Consequently, the conventional master equation is not accurate even in the limit of weak anharmonicity.

Due to this observation, in~\cite{supp} we derive the proper Markovian master equation starting from the first principles, following Refs.~\cite{Hornberger_2009} and~\cite{Walls_2008}. The master equation reads
\begin{align}
\dot{\hat{\rho}} =& - i \omega[\hat{a}^{\dagger}\hat{a},\hat{\rho}(t)] - i \eta[(\hat{a}^{\dagger})^2\hat{a}^2,\hat{\rho}(t)] + \frac{\kappa}{2}\sum_{k=0}^{\infty} (k+1) \nonumber \\
&\times\Big[(\bar{n}_{k+1}+1)\big(2\rho_{k+1,k+1}|k\rangle\langle k| - \{|k+1\rangle\langle k+1|,\hat{\rho}\}\big) \nonumber \\
&\,\,\,\,+\bar{n}_{k+1}\big(2\rho_{k,k}|k+1\rangle\langle k+1| - \{|k\rangle\langle k|,\hat{\rho}\}\big)\Big], \label{eq:203}
\end{align}
where $\bar{n}_k = [ e^{\hbar(\omega+2k\eta)/(k_{\text{B}}T_{\text{E}})} - 1 ]^{-1}$, $\rho_{k,k}=\langle k|\hat{\rho}|k\rangle$, and $|k\rangle \langle k| = \frac{1}{k!}\sum_{p=0}^{\infty} \frac{(-1)^p}{p!}(\hat{a}^{\dagger})^{k+p}\hat{a}^{k+p}$. A similar but less explicit form of the master equation of the anharmonic oscillator is given in~\cite{Alicki_1989}. In~\cite{supp}, we show analytically that this master equation cures the inconsistency presented above in the limit of small $T_{\text{E}}$.

\emph{Example: network of coupled driven dissipative anharmonic oscillators.---}Let us consider a network of $N$ bilinearly coupled, classically driven anharmonic oscillators. Adopting the rotating-wave approximation, the Hamiltonian of the system in the Schr\"odinger picture reads
\begin{align}
\hat{H} / \hbar =& \sum_{j=1}^{N} \Bigg[ \omega_{\text{q},j} \hat{a}_j^{\dagger}\hat{a}_j + \eta_j \big( \hat{a}_j^{\dagger} \big)^2 \hat{a}_j^2 + i \Omega_j \hat{a}_j e^{+i \omega_{\text{d},j}t} + \text{H.c.} \nonumber \\
&+ \sum_{j\neq j'=1}^N J_{j,j'} \hat{a}_j \hat{a}_{j'}^{\dagger} \Bigg], \label{eq:301}
\end{align}
where $\omega_{\text{d},j}$ is the drive frequency and $\Omega_j$ is the complex drive amplitude of the anharmonic oscillator~$j$, H.c.~denotes the Hermitian conjugate of the previous term, and $J_{j,j'}=J_{j',j}$ is the coupling constant between the oscillators $j$ and $j'$.

In the numerical simulations below, we approximate the Hamiltonian to be constant during each small time step. Thus, we can simply replace the Hamiltonian of the system transformed into the Heisenberg picture in the adjoint master equation~(\ref{eq:03}) by the above Schr\"odinger picture Hamiltonian.
Moreover, we model the dissipation in each oscillator with the dissipator presented in Eq.~(\ref{eq:203}), and for simplicity, assume $T_{\text{E}}=0$ and choose the elementary operators as $\hat{o}_{1}=\hat{a}^{\dagger}$ and $\hat{o}_{2}=\hat{a}$. Finally, we define $G(x_1,y_1,\ldots,x_N,y_N;t)=F(x_1,y_1,\ldots,x_N,y_N;t)\prod_{j=1}^N e^{x_j y_j}$ to obtain the following dynamic equation for the transformed generating function
\begin{align}
\dot{G} =&\, \bigg( K_{\text{x}} + K_{\text{y}} + \sum_{j=1}^N \kappa_j P_j \bigg) G, \label{eq:302}
\end{align}
where
\begin{subequations}
\begin{align}
K_{\text{x}} =&\, \sum_{j=1}^{N} \bigg[ \bigg( i\omega_{\text{q},j} - \frac{\kappa_j}{2} + i\sum_{j\neq j'=1}^{N} J_{j,j'} \bigg) x_j \partial_{x_j} \nonumber \\
&\hspace{-2mm}+ i\eta_j x_j^2 \partial_{x_j}^2 - \Omega_j e^{+i \omega_{\text{d},j}t} \partial_{x_j} + \Omega_j^* e^{-i \omega_{\text{d},j}t} x_j \bigg] , \label{eq:302_2a} \\
K_{\text{y}} =&\, \sum_{j=1}^{N} \bigg[ \bigg( -i\omega_{\text{q},j} - \frac{\kappa_j}{2} - i\sum_{j\neq j'=1}^{N} J_{j,j'} \bigg) y_j \partial_{y_j} \nonumber \\
&\hspace{-2mm}- i\eta_j y_j^2 \partial_{y_j}^2 + \Omega_j e^{+i \omega_{\text{d},j}t} y_j - \Omega_j^* e^{-i \omega_{\text{d},j}t} \partial_{y_j} \bigg], \label{eq:302_2b} \\
P_jG &= x_j y_j \nonumber \\
&\hspace{-1mm} \times \sum_{k_1,l_1,\ldots,k_N,l_N}\delta_{k_j,l_j}g_{k_1,l_1,\ldots,k_N,l_N}x_1^{k_1}y_1^{l_1}\ldots x_N^{k_N}y_N^{l_N} \label{eq:302_2c},
\end{align}
\end{subequations}
where we have used the short-hand notation $G \equiv G(x_1,y_1,\ldots,x_N,y_N;t)$. Similarly to the previous sections, we introduce three new complex-valued functions, which yields
\begin{subequations}
\begin{align}
\dot{G}_{\textbf{x}} =&\, \big( K_{\text{x}} + C \big) G_{\textbf{x}}, \label{eq:303a} \\
\dot{G}_{\textbf{y}} =&\, \big( K_{\text{y}} - C \big) G_{\textbf{y}}, \label{eq:303b} \\
\dot{R} =&\, \bigg( K_{\text{x}} + K_{\text{y}} + \sum_{j=1}^N \kappa_j P_j \bigg) R + \sum_{j=1}^N \kappa_j P_j G_{\textbf{x}} G_{\textbf{y}}. \label{eq:303c}
\end{align}
\end{subequations}
To solve the dynamics of the operator under interest, Eqs.~(\ref{eq:303a})--(\ref{eq:303c}) are solved with several initial conditions, $G^{(p_1,q_1,\ldots,p_N,q_N)}(x_1,y_1,\ldots,x_N,y_N;0)=\prod_{j=1}^N x_j^{p_j}y_j^{q_j}$. The separation constant $C$ is chosen such that it provides sufficient numerical stability. Here, we use the constant that minimizes the energy of the normalized derivatives, $|\dot{G}_{\textbf{x}}/G_{\textbf{x}}|^2 + |\dot{G}_{\textbf{y}}/G_{\textbf{y}}|^2$, of the uncoupled system at $t=0$. Such a constant depends on the initial conditions and reads
\begin{align}
C=&\,-\frac{i}{2}\sum_{j=1}^N \big[ (\omega_{\text{q},j} - \eta_j )(p_j+q_j) + \eta_j \big( p_j^2 + q_j^2 \big) \big]. \label{eq:304}
\end{align}

\begin{table}[t!]
\renewcommand*{\arraystretch}{1.7}
\caption{Physical parameters used in the numerical simulations. \vspace{2mm} \label{tab:01}}
\begin{tabular}{ c | c  c  c  c  c  c c }
Parameter & $\omega_{\text{q},j}$ & $\omega_{\text{d},j}$ & $\eta_j$ & $\Omega_1$ & $\Omega_{j>1}$ & $\kappa_j$ & $J_{j,j'}$ \\
\hline
Value ($2\pi\times$MHz) & $3500$ & $3500$ & $250$ & $125$ & $0$ & $0.02/2\pi$ & $10$ \\
\end{tabular}
\end{table}

\begin{figure}[b!]
  \centering
  \includegraphics[width = \columnwidth]{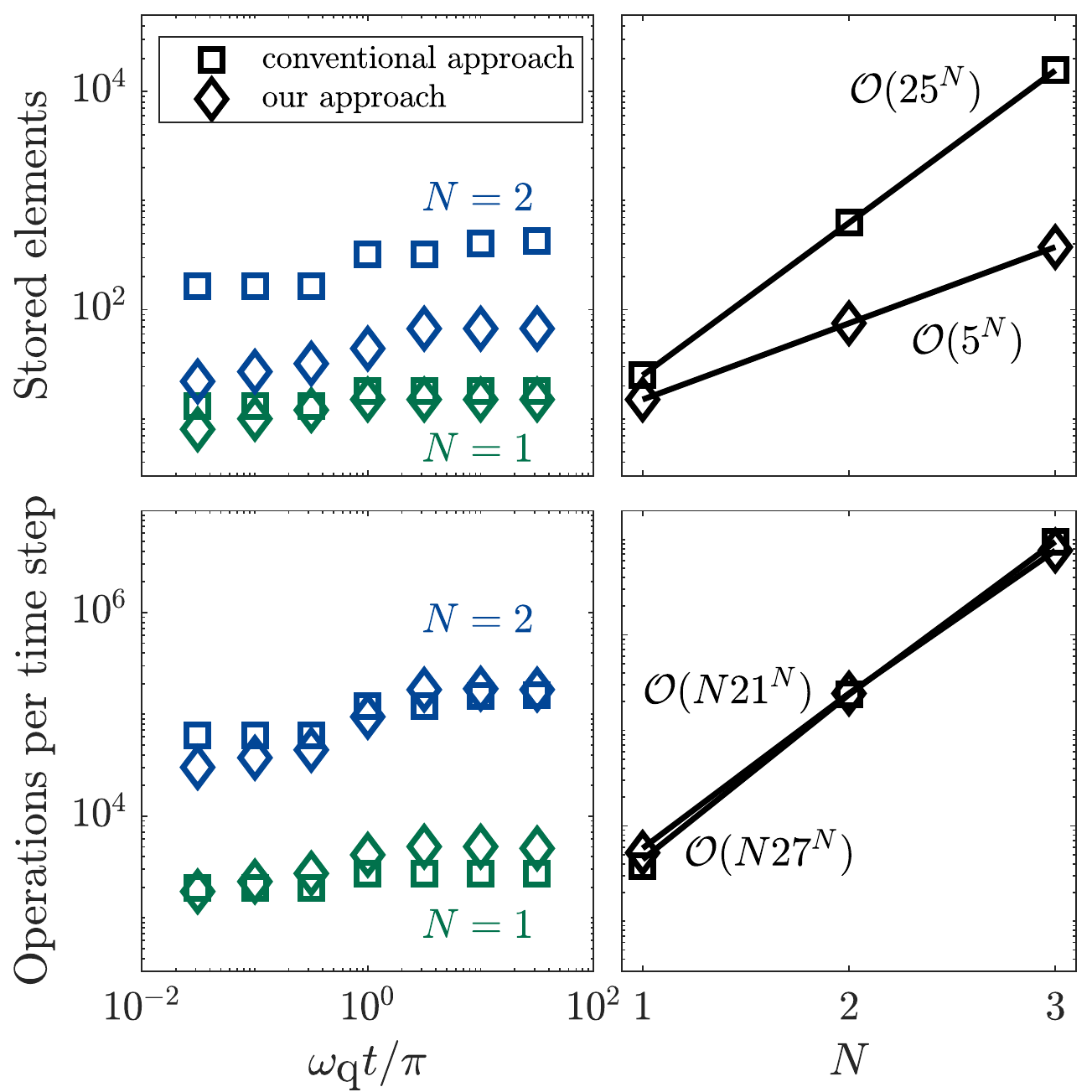} 
  \caption{\label{fig:2} Largest number of stored elements during the simulation (top panels) and the number of computational elementary operations per time step (bottom panels) as functions of scaled time (left panels) and the number of subsystems (right panels). The scaled time used in the right panels is $\omega_{\text{q}} t/\pi = 10$. The squares correspond to the conventional approach and the diamonds to our approach. The green symbols correspond to the number of oscillators $N=1$ and the blue symbols to $N=2$.}
\end{figure}

We implement numerical solvers based on our approach and on the conventional approach.
The physical simulation parameters are chosen to correspond to a typical superconducting quantum processor, and are given in Table~\ref{tab:01}. The operator for which we solve the dynamics is $\hat{A}(0)=\tens{j=1}{N}\hat{a}_{j}$. For both methods, the numerical integration is implemented using the fourth-order Runge--Kutta method. By tuning the threshold for truncating the representation of the operator, we force the relative errors $\epsilon_{\text{r}}$ to be approximately constant, $100\text{ ppm}$, for all the simulations. The relative error is defined as $\epsilon_{\text{r}} = \| M-M_{\text{ref}}\|_2/\| M_{\text{ref}}\|_2$, where $\| *\|_2$ is the element-wise $L^2$ norm, $M$ is the representation of the operator in the number basis truncated to three least energetic states, and $M_{\text{ref}}$ is that given by the reference method. The reference method is the conventional method executed by storing a larger representation during the simulation, and by using half the time step.

The results in Fig.~\ref{fig:2} show that both the number of the stored elements and the number of operations increase with increasing time of the evolution of the system for both of the methods. Furthermore, we find that increasing $\kappa_j t$ to values for which the residual function $R$ cannot be approximated as a zero function, it is nevertheless well-approximated by a function of $N$ variables only, namely, $\{ x_j y_j\}_{j=1}^N$. The number of stored elements of the conventional approach scales as the square of that required by our approach as a function of the number of subsystems. This is expected since our approach only requires storing $N$-variable functions during the simulation whereas the conventional approach requires storing a $2N$-variable function. Finally, the number of operations scales similarly for both of the methods as a function of $N$. This is expected since increasing $N$ increases the number of generating functions, the dynamics of which are to be solved for our method.

\emph{Conclusions.---}We have introduced a numerically efficient approach to solve the dynamics of certain operators of weakly open bosonic systems.
The approach is based on expanding the operators of the system in a problem-specific operator order, mapping the adjoint master equation into a dynamic equation for a generating function of this expansion, applying a transformation to the dynamic equation, and solving it by numerical integration.
By considering a network of classically driven damped anharmonic oscillators, we demonstrate that our approach reduces the memory consumption compared to the conventional method quadratically.
Moreover, we observe that the computation times of the methods are comparable. However, parallelization of the computationally heavy steps of our method enables reaching a speed-up compared with the conventional approach. In the future, the generality of the proposed approach enables applying it to a range of weakly open bosonic systems under topical interest. Moreover, the technique may be further generalized by considering expansions in more general operator orders and function bases. This may broaden the applicability of the proposed technique even further.

\begin{acknowledgments}
This research was financially supported by the European Research Council under Grant No.~681311 (QUESS); by the Academy of Finland under its Centers of Excellence Program Grant No.~336810; and by the Vilho, Yrj\"{o} and Kalle V\"{a}is\"{a}l\"{a} Foundation of the Finnish Academy of Science and Letters and Finnish Cultural Foundation. This work was also supported by Leading Initiative for Excellent Young Researchers MEXT Japan and JST presto (Grant No.~JPMJPR1919) Japan.\\
\end{acknowledgments}

\emph{Competing interests.---}The authors declare that IQM has filed a patent application regarding the simulation method in February, 2022.

\bibliographystyle{apsrev4-1}
\bibliography{main}

\end{document}


\preprint{AIP/123-QED}

\title{Supplemental Materials: Efficient classical simulation of open\\ bosonic quantum systems}

\author{Akseli M\"akinen}
\affiliation{IQM, Keilaranta 19, FI-02150 Espoo, Finland}
\affiliation{QCD Labs, QTF Centre of Excellence, Department of Applied Physics, Aalto University, P.O.~Box 13500, FI-00076 Aalto, Finland}
\affiliation{Research Center for Emerging Computing Technologies (RCECT), National Institute of Advanced Industrial Science and Technology (AIST), 1-1-1 Umezono, Tsukuba, Ibaraki 305-8568, Japan}
\author{Joni Ikonen}
\affiliation{IQM, Keilaranta 19, FI-02150 Espoo, Finland}
\affiliation{QCD Labs, QTF Centre of Excellence, Department of Applied Physics, Aalto University, P.O.~Box 13500, FI-00076 Aalto, Finland}
\author{Takaaki Aoki}
\affiliation{Research Center for Emerging Computing Technologies (RCECT), National Institute of Advanced Industrial Science and Technology (AIST), 1-1-1 Umezono, Tsukuba, Ibaraki 305-8568, Japan}
\affiliation{Department of Physics, The University of Tokyo, 5-1-5 Kashiwanoha, Kashiwa, Chiba 277-8574, Japan}
\author{Jani Tuorila}
\affiliation{IQM, Keilaranta 19, FI-02150 Espoo, Finland}
\author{Yuichiro Matsuzaki}
\affiliation{Research Center for Emerging Computing Technologies (RCECT), National Institute of Advanced Industrial Science and Technology (AIST), 1-1-1 Umezono, Tsukuba, Ibaraki 305-8568, Japan}
\affiliation{NEC-AIST Quantum Technology Cooperative Research Laboratory, National Institute of Advanced Industrial Science and Technology (AIST), Tsukuba, Ibaraki 305-8568, Japan}
\author{Mikko M\"ott\"onen}
\affiliation{QCD Labs, QTF Centre of Excellence, Department of Applied Physics, Aalto University, P.O.~Box 13500, FI-00076 Aalto, Finland}
\affiliation{VTT Technical Research Centre of Finland Ltd., QTF Center of Excellence, P.O. Box 1000, FI-02044 VTT, Finland}

\date{\today}
\maketitle

\widetext
\setcounter{equation}{0}
\setcounter{figure}{0}
\setcounter{table}{0}
\makeatletter
\renewcommand{\theequation}{S\arabic{equation}}
\renewcommand{\thefigure}{S\arabic{figure}}

\section{Dynamic equation for the generating function of an arbitrary operator of an arbitrary multimode bosonic system}
\label{sec:der}

In this section, we derive the dynamic equation for the generating function of an operator of a general bosonic system consisting of $N$ discrete modes. For simplicity, we assume here that the system is closed. However, we demonstrate a derivation of the part of the dynamic equation related to the Lindblad part of the adjoint master equation for the anharmonic oscillator in Eqs.~(\ref{eq:312})--(\ref{eq:314}). Similarly to the main text, we define for each mode $j$ two elementary operators $\hat{o}_{j,1}$ and $\hat{o}_{j,2}$ that obey the commutation relation $[ \hat{o}_{j,2}, \hat{o}_{j,1} ]=r_j\hat{I}$, and further note that the operators between different modes commute. In terms of these operators, the Hamiltonian of the system in the Heisenberg picture can be expanded as 
\begin{align}
\hat{H}(t) =&\, \hbar \sum_{(\mathbf{u},\mathbf{v}) \in \mathbb{N}_0^{2N}} h_{\mathbf{u},\mathbf{v}}(t) \tens{j=1}{N} \hat{o}_{j,1}^{u_j} \hat{o}_{j,2}^{v_j}, \label{eq:101}
\end{align}
where $\mathbf{u}=\{ u_1,\ldots,u_N \}$ and $\mathbf{v}=\{ v_1,\ldots,v_N \}$, and an arbitrary operator of the system as
\begin{align}
\hat{A}(t) =& \sum_{(\mathbf{k},\mathbf{l}) \in \mathbb{N}_0^{2N}} f_{\mathbf{k},\mathbf{l}}(t) \tens{j=1}{N} \hat{o}_{j,1}^{k_j} \hat{o}_{j,2}^{l_j}. \label{eq:102}
\end{align}
In the following, the summations range from $0$ to $\infty$ for each index unless otherwise stated. Assuming that the operator does not depend explicitly on time, it evolves according to $\dot{\hat{A}}(t) = \,\frac{i}{\hbar}\big[\hat{H}(t),\hat{A}(t)\big]$, which leads to
\begingroup
\allowdisplaybreaks
\begin{align}
\sum_{\mathbf{k},\mathbf{l}}\dot{f}_{\mathbf{k},\mathbf{l}}(t)\tens{j=1}{N}\hat{o}_{j,1}^{k_{j}}\hat{o}_{j,2}^{l_{j}} =& \,i\Bigg[\sum_{\mathbf{u},\mathbf{v}}h_{\mathbf{u},\mathbf{v}}(t)\tens{j=1}{N}\hat{o}_{j,1}^{u_{j}}\hat{o}_{j,2}^{v_{j}},\sum_{\mathbf{k},\mathbf{l}}f_{\mathbf{k},\mathbf{l}}(t)\tens{j=1}{N}\hat{o}_{j,1}^{k_{j}}\hat{o}_{j,2}^{l_{j}}\Bigg] \nonumber \\
=&\,i\sum_{\mathbf{u},\mathbf{v}}h_{\mathbf{u},\mathbf{v}}(t)\sum_{\mathbf{k},\mathbf{l}}f_{\mathbf{k},\mathbf{l}}(t)\Bigg(\tens{j=1}{N}\hat{o}_{j,1}^{u_{j}}\hat{o}_{j,2}^{v_{j}}\hat{o}_{j,1}^{k_{j}}\hat{o}_{j,2}^{l_{j}}-\tens{j=1}{N}\hat{o}_{j,1}^{k_{j}}\hat{o}_{j,2}^{l_{j}}\hat{o}_{j,1}^{u_{j}}\hat{o}_{j,2}^{v_{j}}\Bigg) \nonumber \\
=&\,i\sum_{\mathbf{u},\mathbf{v}}h_{\mathbf{u},\mathbf{v}}(t)\sum_{\mathbf{k},\mathbf{l}}f_{\mathbf{k},\mathbf{l}}(t)\Bigg[\tens{j=1}{N}\hat{o}_{j,1}^{u_{j}}\sum_{s_{j}=0}^{\min(v_{j},k_{j})}r_{j}^{s_{j}}s_{j}!\binom{v_{j}}{s_{j}}\binom{k_{j}}{s_{j}}\hat{o}_{j,1}^{k_{j}-s_{j}}\hat{o}_{j,2}^{v_{j}-s_{j}}\hat{o}_{j,2}^{l_{j}}\nonumber\\
&\,-\tens{j=1}{N}\hat{o}_{j,1}^{k_{j}}\sum_{s_{j}=0}^{\min(l_{j},u_{j})}r_{j}^{s_{j}}s_{j}!\binom{l_{j}}{s_{j}}\binom{u_{j}}{s_{j}}\hat{o}_{j,1}^{u_{j}-s_{j}}\hat{o}_{j,2}^{l_{j}-s_{j}}\hat{o}_{j,2}^{v_{j}}\Bigg]\nonumber \displaybreak \\
=& \,i\sum_{\mathbf{u},\mathbf{v}}h_{\mathbf{u},\mathbf{v}}(t)\sum_{\mathbf{k},\mathbf{l}}f_{\mathbf{k},\mathbf{l}}(t)\Bigg[\tens{j=1}{N}\sum_{s_{j}=0}^{\min(v_{j},k_{j})}r_{j}^{s_{j}}s_{j}!\binom{v_{j}}{s_{j}}\binom{k_{j}}{s_{j}}\hat{o}_{j,1}^{k_{j}+u_{j}-s_{j}}\hat{o}_{j,2}^{l_{j}+v_{j}-s_{j}}\nonumber \\
&\,-\tens{j=1}{N}\sum_{s_{j}=0}^{\min(l_{j},u_{j})}r_{j}^{s_{j}}s_{j}!\binom{l_{j}}{s_{j}}\binom{u_{j}}{s_{j}}\hat{o}_{j,1}^{k_{j}+u_{j}-s_{j}}\hat{o}_{j,2}^{l_{j}+v_{j}-s_{j}}\Bigg], \label{eq:103}
\end{align}
\endgroup
where in the second step we have used the commutation relations of operator monomials~\cite{Pain_2012}.
By defining the binomial coefficients as $\binom{n}{m}=\frac{\Gamma(n+1)}{\Gamma(m+1)\Gamma(n-m+1)}$ we can extend the summations over $s_j$'s to infinity in order to simplify the reorganization of the summations. Consequently,
\begin{align}
\sum_{\mathbf{k},\mathbf{l}}\dot{f}_{\mathbf{k},\mathbf{l}}(t)\tens{j=1}{N}\hat{o}_{j,1}^{k_{j}}\hat{o}_{j,2}^{l_{j}}=&\,i\sum_{\mathbf{u},\mathbf{v}}h_{\mathbf{u},\mathbf{v}}(t)\sum_{\mathbf{k},\mathbf{l}}f_{\mathbf{k},\mathbf{l}}(t)\tens{j=1}{N}\sum_{s_{j}=0}^{\infty}r_{j}^{s_{j}}s_{j}!\binom{v_{j}}{s_{j}}\binom{k_{j}}{s_{j}}\hat{o}_{j,1}^{k_{j}+u_{j}-s_{j}}\hat{o}_{j,2}^{l_{j}+v_{j}-s_{j}}\nonumber\\
&\,-i\sum_{\mathbf{u},\mathbf{v}}h_{\mathbf{u},\mathbf{v}}(t)\sum_{\mathbf{k},\mathbf{l}}f_{\mathbf{k},\mathbf{l}}(t)\tens{j=1}{N}\sum_{s_{j}=0}^{\infty}r_{j}^{s_{j}}s_{j}!\binom{l_{j}}{s_{j}}\binom{u_{j}}{s_{j}}\hat{o}_{j,1}^{k_{j}+u_{j}-s_{j}}\hat{o}_{j,2}^{l_{j}+v_{j}-s_{j}}\nonumber\\
=&\,i\sum_{\mathbf{u},\mathbf{v}}h_{\mathbf{u},\mathbf{v}}(t)\tens{j=1}{N}\sum_{\mathbf{s}}\sum_{\mathbf{k},\mathbf{l}}f_{\mathbf{k},\mathbf{l}}(t)r_{j}^{s_{j}}s_{j}!\binom{v_{j}}{s_{j}}\binom{k_{j}}{s_{j}}\hat{o}_{j,1}^{k_{j}+u_{j}-s_{j}}\hat{o}_{j,2}^{l_{j}+v_{j}-s_{j}}\nonumber\\
&\,-i\sum_{\mathbf{u},\mathbf{v}}h_{\mathbf{u},\mathbf{v}}(t)\tens{j=1}{N}\sum_{\mathbf{s}}\sum_{\mathbf{k},\mathbf{l}}f_{\mathbf{k},\mathbf{l}}(t)r_{j}^{s_{j}}s_{j}!\binom{l_{j}}{s_{j}}\binom{u_{j}}{s_{j}}\hat{o}_{j,1}^{k_{j}+u_{j}-s_{j}}\hat{o}_{j,2}^{l_{j}+v_{j}-s_{j}}\nonumber\\
=&\,i\sum_{\mathbf{u},\mathbf{v}}h_{\mathbf{u},\mathbf{v}}(t)\tens{j=1}{N}\sum_{\mathbf{s}}\sum_{\mathbf{k},\mathbf{l}}f_{\mathbf{k}-\mathbf{u}+\mathbf{s},\mathbf{l}-\mathbf{v}+\mathbf{s}}(t)r_{j}^{s_{j}}s_{j}!\binom{v_{j}}{s_{j}}\binom{k_{j}-u_{j}+s_{j}}{s_{j}}\hat{o}_{j,1}^{k_{j}}\hat{o}_{j,2}^{l_{j}}\nonumber\\
&\,-i\sum_{\mathbf{u},\mathbf{v}}h_{\mathbf{u},\mathbf{v}}(t)\tens{j=1}{N}\sum_{\mathbf{s}}\sum_{\mathbf{k},\mathbf{l}}f_{\mathbf{k}-\mathbf{u}+\mathbf{s},\mathbf{l}-\mathbf{v}+\mathbf{s}}(t)r_{j}^{s_{j}}s_{j}!\binom{l_{j}-v_{j}+s_{j}}{s_{j}}\binom{u_{j}}{s_{j}}\hat{o}_{j,1}^{k_{j}}\hat{o}_{j,2}^{l_{j}}. \label{eq:104}
\end{align}
Using the completeness of the operator monomials we directly obtain
\begin{align}
\dot{f}_{\mathbf{k},\mathbf{l}}(t)=&\,i\sum_{\mathbf{u},\mathbf{v}}h_{\mathbf{u},\mathbf{v}}(t)\sum_{\mathbf{s}}f_{\mathbf{k}-\mathbf{u}+\mathbf{s},\mathbf{l}-\mathbf{v}+\mathbf{s}}(t)r_{j}^{s_{j}}s_{j}!\binom{v_{j}}{s_{j}}\binom{k_{j}-u_{j}+s_{j}}{s_{j}}\nonumber\\
&\,-i\sum_{\mathbf{u},\mathbf{v}}h_{\mathbf{u},\mathbf{v}}(t)\sum_{\mathbf{s}}f_{\mathbf{k}-\mathbf{u}+\mathbf{s},\mathbf{l}-\mathbf{v}+\mathbf{s}}(t)r_{j}^{s_{j}}s_{j}!\binom{l_{j}-v_{j}+s_{j}}{s_{j}}\binom{u_{j}}{s_{j}}. \label{eq:105}
\end{align}
Multiplying the both sides by $\prod_{j=1}^{N}x_j^{k_j}y_j^{l_j}$ and performing the summation over the indices $k_j$ and $l_j$ gives
\begin{align}
\sum_{\mathbf{k},\mathbf{l}}\prod_{j=1}^{N}x_j^{k_j}y_j^{l_j}\dot{f}_{\mathbf{k},\mathbf{l}}(t)=&\,i\sum_{\mathbf{k},\mathbf{l}}\prod_{j=1}^{N}x_j^{k_j}y_j^{l_j}\sum_{\mathbf{u},\mathbf{v}}h_{\mathbf{u},\mathbf{v}}(t)\sum_{\mathbf{s}}f_{\mathbf{k}-\mathbf{u}+\mathbf{s},\mathbf{l}-\mathbf{v}+\mathbf{s}}(t)r_{j}^{s_{j}}s_{j}!\binom{v_{j}}{s_{j}}\binom{k_{j}-u_{j}+s_{j}}{s_{j}}\nonumber\\
&\,-i\sum_{\mathbf{k},\mathbf{l}}\prod_{j=1}^{N}x_j^{k_j}y_j^{l_j}\sum_{\mathbf{u},\mathbf{v}}h_{\mathbf{u},\mathbf{v}}(t)\sum_{\mathbf{s}}f_{\mathbf{k}-\mathbf{u}+\mathbf{s},\mathbf{l}-\mathbf{v}+\mathbf{s}}(t)r_{j}^{s_{j}}s_{j}!\binom{l_{j}-v_{j}+s_{j}}{s_{j}}\binom{u_{j}}{s_{j}}. \label{eq:106}
\end{align}
Defining a generating function related to the operator $\hat{A}(t)$ in the chosen operator order as $F(\mathbf{x},\mathbf{y};t)=\sum_{\mathbf{k},\mathbf{l}}f_{\mathbf{k},\mathbf{l}}(t)\prod_{j=1}^{N}x_j^{k_j}y_j^{l_j}$ finally gives us
\begingroup
\allowdisplaybreaks
\begin{align}
\dot{F}(\mathbf{x},\mathbf{y};t)=&\,i\sum_{\mathbf{u},\mathbf{v}}h_{\mathbf{u},\mathbf{v}}(t)\sum_{\mathbf{s},\mathbf{k},\mathbf{l}}f_{\mathbf{k}-\mathbf{u}+\mathbf{s},\mathbf{l}-\mathbf{v}+\mathbf{s}}(t)\prod_{j=1}^{N}x_{j}^{k_{j}}y_{j}^{l_{j}}r_{j}^{s_{j}}\binom{v_{j}}{s_{j}}\frac{(k_{j}-u_{j}+s_{j})!}{(k_{j}-u_{j})!}\nonumber \\
&\,-i\sum_{\mathbf{u},\mathbf{v}}h_{\mathbf{u},\mathbf{v}}(t)\sum_{\mathbf{s},\mathbf{k},\mathbf{l}}f_{\mathbf{k}-\mathbf{u}+\mathbf{s},\mathbf{l}-\mathbf{v}+\mathbf{s}}(t)\prod_{j=1}^{N}x_{j}^{k_{j}}y_{j}^{l_{j}}r_{j}^{s_{j}}\frac{(l_{j}-v_{j}+s_{j})!}{(l_{j}-v_{j})!}\binom{u_{j}}{s_{j}}\nonumber \\
=&\,i\sum_{\mathbf{u},\mathbf{v}}h_{\mathbf{u},\mathbf{v}}(t)\sum_{\mathbf{s},\mathbf{k},\mathbf{l}}f_{\mathbf{k},\mathbf{l}}(t)\prod_{j=1}^{N}x_{j}^{k_{j}+u_{j}-s_{j}}y_{j}^{l_{j}+v_{j}-s_{j}}r_{j}^{s_{j}}\binom{v_{j}}{s_{j}}\frac{k_{j}!}{(k_{j}-s_{j})!}\nonumber \\
&\,-i\sum_{\mathbf{u},\mathbf{v}}h_{\mathbf{u},\mathbf{v}}(t)\sum_{\mathbf{s},\mathbf{k},\mathbf{l}}f_{\mathbf{k},\mathbf{l}}(t)\prod_{j=1}^{N}x_{j}^{k_{j}+u_{j}-s_{j}}y_{j}^{l_{j}+v_{j}-s_{j}}r_{j}^{s_{j}}\frac{l_{j}!}{(l_{j}-s_{j})!}\binom{u_{j}}{s_{j}}\nonumber \\
=&\,i\sum_{\mathbf{u},\mathbf{v}}h_{\mathbf{u},\mathbf{v}}(t)\sum_{\mathbf{s},\mathbf{k},\mathbf{l}}f_{\mathbf{k},\mathbf{l}}(t)\prod_{j=1}^{N}x_{j}^{u_{j}}y_{j}^{v_{j}-s_{j}}r_{j}^{s_{j}}\binom{v_{j}}{s_{j}}\frac{k_{j}!}{(k_{j}-s_{j})!}x_{j}^{k_{j}-s_{j}}y_{j}^{l_{j}}\nonumber \\
&\,-i\sum_{\mathbf{u},\mathbf{v}}h_{\mathbf{u},\mathbf{v}}(t)\sum_{\mathbf{s},\mathbf{k},\mathbf{l}}f_{\mathbf{k},\mathbf{l}}(t)\prod_{j=1}^{N}x_{j}^{u_{j}-s_{j}}y_{j}^{v_{j}}r_{j}^{s_{j}}\binom{u_{j}}{s_{j}}\frac{l_{j}!}{(l_{j}-s_{j})!}x_{j}^{k_{j}}y_{j}^{l_{j}-s_{j}}\nonumber \\
=&\,i\sum_{\mathbf{u},\mathbf{v}}h_{\mathbf{u},\mathbf{v}}(t)\sum_{\mathbf{k},\mathbf{l}}f_{\mathbf{k},\mathbf{l}}(t)\prod_{j=1}^{N}x_{j}^{u_{j}}\sum_{s_j}\binom{v_{j}}{s_{j}}y_{j}^{v_{j}-s_{j}}r_{j}^{s_{j}}\partial_{x_{j}}^{s_{j}}x_{j}^{k_{j}}y_{j}^{l_{j}}\nonumber \\
&\,-i\sum_{\mathbf{u},\mathbf{v}}h_{\mathbf{u},\mathbf{v}}(t)\sum_{\mathbf{k},\mathbf{l}}f_{\mathbf{k},\mathbf{l}}(t)\prod_{j=1}^{N}y_{j}^{v_{j}}\sum_{s_j}\binom{u_{j}}{s_{j}}x_{j}^{u_{j}-s_{j}}r_{j}^{s_{j}}\partial_{y_{j}}^{s_{j}}x_{j}^{k_{j}}y_{j}^{l_{j}}\nonumber \\
=&\,i\sum_{\mathbf{u},\mathbf{v}}h_{\mathbf{u},\mathbf{v}}(t)\sum_{\mathbf{k},\mathbf{l}}f_{\mathbf{k},\mathbf{l}}(t)\prod_{j=1}^{N}x_{j}^{u_{j}}\big(y_{j}+r_{j}\partial_{x_{j}}\big)^{v_{j}}x_{j}^{k_{j}}y_{j}^{l_{j}}-i\sum_{\mathbf{u},\mathbf{v}}h_{\mathbf{u},\mathbf{v}}(t)\sum_{\mathbf{k},\mathbf{l}}f_{\mathbf{k},\mathbf{l}}(t)\prod_{j=1}^{N}y_{j}^{v_{j}}\big(x_{j}+r_{j}\partial_{y_{j}}\big)^{u_{j}}x_{j}^{k_{j}}y_{j}^{l_{j}}\nonumber \\
=&\,i\sum_{\mathbf{u},\mathbf{v}}h_{\mathbf{u},\mathbf{v}}(t)\prod_{j=1}^{N}x_{j}^{u_{j}}\big(y_{j}+r_{j}\partial_{x_{j}}\big)^{v_{j}}\sum_{\mathbf{k},\mathbf{l}}f_{\mathbf{k},\mathbf{l}}(t)x_{j}^{k_{j}}y_{j}^{l_{j}}-i\sum_{\mathbf{u},\mathbf{v}}h_{\mathbf{u},\mathbf{v}}(t)\prod_{j=1}^{N}y_{j}^{v_{j}}\big(x_{j}+r_{j}\partial_{y_{j}}\big)^{u_{j}}\sum_{\mathbf{k},\mathbf{l}}f_{\mathbf{k},\mathbf{l}}(t)x_{j}^{k_{j}}y_{j}^{l_{j}}\nonumber \\
\Rightarrow \dot{F}(\mathbf{x},\mathbf{y};t)=&\,i\big[:H\big(\mathbf{x},\mathbf{r}\circ\partial_{\mathbf{x}}+\mathbf{y}; t\big):-:H\big(\mathbf{x}+\mathbf{r}\circ\partial_{\mathbf{y}},\mathbf{y}; t\big):\big]F(\mathbf{x},\mathbf{y};t), \label{eq:107}
\end{align}
\endgroup
where $\mathbf{r}=(r_1,\ldots,r_N)^{\text{T}}$, T denotes the transpose, $:*:$ is an operator that orders the variables to the left with respect to the differentiation operators, $\circ$ denotes the Hadamard product, and $H(\mathbf{x},\mathbf{y};t)=\sum_{\mathbf{u},\mathbf{v}}h_{\mathbf{u},\mathbf{v}}(t)\prod_{j=1}^{N}x_j^{u_j}y_j^{v_j}$. This concludes the derivation. Note that for single-mode systems this equation reduces to the first two lines of Eq.~(4) of the main text.

\section{The anharmonic oscillator}
\label{sec:mas}

\textbf{Master equation}---In this section, we derive the master equation for the anharmonic oscillator starting from the first principles, following Refs.~\cite{Hornberger_2009, Walls_2008}. In the limit of weak interaction between the system under interest and its environment, the microscopic Hamiltonian reads~\cite{Milburn_1986}
\begin{align}
\hat{H}_{\text{tot}}=&\,\hat{H}+\hat{H}_{\text{E}}+\hat{H}_{\text{I}}, \label{eq:301}
\end{align}
where
\begin{subequations}
\begin{align}
\hat{H}=&\,\hbar\omega\hat{a}^{\dagger}\hat{a}+\hbar\eta\big(\hat{a}^{\dagger}\big)^{2}\hat{a}^{2}, \label{eq:302a} \\
\hat{H}_{\text{E}}=&\,\hbar\sum_{k}\omega_{k}\hat{b}_{k}^{\dagger}\hat{b}_{k}, \label{eq:302b} \\
\hat{H}_{\text{I}}=&\,\hbar\sum_{k}\textsl{g}_{k}\big(\hat{a}+\hat{a}^{\dagger}\big)\big(\hat{b}_{k}+\hat{b}_{k}^{\dagger}\big) \label{eq:302c} \\
=&\,\hat{A}\sum_{k}\hat{B}_{k}, \label{eq:302d}
\end{align}
\end{subequations}
are the Hamiltonians of the anharmonic oscillator, the environment, and the bilinear coupling between the system and the environment. Here, we have defined $\hat{A}=\hat{a}+\hat{a}^{\dagger}$ and $\hat{B}_{k}=\hbar \textsl{g}_{k}\big(\hat{b}_{k}+\hat{b}_{k}^{\dagger}\big)$. In the energy eigenbasis of the system under interest, we have
\begingroup
\allowdisplaybreaks
\begin{align}
\hat{A}(\xi)=&\,\sum_{E',E} \mathbbm{1}(E'-E=\hbar\xi) \langle E|\hat{A}|E'\rangle|E\rangle\langle E'| \nonumber \\
=&\,\sum_{n,m=0}^{\infty}\mathbbm{1}[\hbar m(\omega-\eta+\eta m)-\hbar n(\omega-\eta+\eta n)=\hbar\xi] \langle n|\big(\hat{a}+\hat{a}^{\dagger}\big)|m\rangle|n\rangle\langle m| \nonumber \\
=&\,\sum_{n,m=0}^{\infty}\mathbbm{1}[\hbar m(\omega-\eta+\eta m)-\hbar n(\omega-\eta+\eta n)=\hbar\xi] \big(\sqrt{m}\langle n|m-1\rangle+\sqrt{m+1}\langle n|m+1\rangle\big)|n\rangle\langle m| \nonumber \\
=&\,\sum_{n=0}^{\infty}\mathbbm{1} \{ \hbar(n+1)[\omega-\eta+\eta(n+1)]-\hbar n(\omega-\eta+\eta n)=\hbar\xi \} \sqrt{n+1}|n\rangle\langle n+1| \nonumber \\
&\,+\sum_{n=1}^{\infty}\mathbbm{1} \{ \hbar(n-1)[\omega-\eta+\eta(n-1)]-\hbar n(\omega-\eta+\eta n)=\hbar\xi \} \sqrt{n}|n\rangle\langle n-1| \nonumber \\
=&\,\sum_{n=0}^{\infty}\big[\mathbbm{1}(\omega+2\eta n=\xi)\sqrt{n+1}|n\rangle\langle n+1|+\mathbbm{1}(\omega+2\eta n=-\xi)\sqrt{n+1}|n+1\rangle\langle n|\big], \label{eq:303}
\end{align}
\endgroup
where $\mathbbm{1}(*)$ is the indicator function. In the interaction picture with respect to $\hat{H}+\hat{H}_{\text{E}}$, $\hat{B}_{k}$ reads
\begin{align}
\tilde{\hat{B}}_{k}(t)=&\,e^{+i (\hat{H}+\hat{H}_{\text{E}} )t/\hbar}\hat{B}_{k}e^{-i (\hat{H}+\hat{H}_{\text{E}} )t/\hbar} \nonumber \\
=&\,\hbar \textsl{g}_{k}e^{-i\omega_{k}t}\hat{b}_{k}+\hbar \textsl{g}_{k}e^{+i\omega_{k}t}\hat{b}_{k}^{\dagger}. \label{eq:304}
\end{align}
Consequently,
\begin{align}
\gamma_{k,l}(\xi)=&\,\frac{1}{\hbar^{2}}\int_{-\infty}^{\infty}\text{d}t\,e^{i\xi t}\big\langle\tilde{\hat{B}}_{k}(t)\hat{B}_{l}(0)\big\rangle \nonumber \\
\approx&\,\int_{-\infty}^{\infty}\text{d}t\,e^{i\xi t}\textsl{g}_{k}\textsl{g}_{l}\Big(e^{-i\omega_{k}t}\big\langle\hat{b}_{k}\hat{b}_{l}^{\dagger}\big\rangle+e^{+i\omega_{k}t}\big\langle\hat{b}_{k}^{\dagger}\hat{b}_{l}\big\rangle\Big), \label{eq:305}
\end{align}
where we have used $\big\langle\hat{b}_{k}\hat{b}_{l}\big\rangle=\big\langle\hat{b}_{k}^{\dagger}\hat{b}_{l}^{\dagger}\big\rangle=0$, which holds for environments consisting of harmonic oscillators in thermal equilibrium. In the interaction picture, the master equation after neglecting the Lamb shift reads
\begin{align}
\partial_{t}\tilde{\hat{\rho}}(t)=&\,\sum_{\xi}\sum_{k,l}\gamma_{k,l}(\xi)\bigg[\hat{A}(\xi)\tilde{\hat{\rho}}(t)\hat{A}^{\dagger}(\xi)-\frac{1}{2}\hat{A}^{\dagger}(\xi)\hat{A}(\xi)\tilde{\hat{\rho}}(t)-\frac{1}{2}\tilde{\hat{\rho}}(t)\hat{A}^{\dagger}(\xi)\hat{A}(\xi)\bigg] \nonumber \\
=&\,\sum_{\xi}\sum_{k,l}\int_{-\infty}^{\infty}\text{d}t\,e^{i\xi t}\textsl{g}_{k}\textsl{g}_{l}\Big(e^{-i\omega_{k}t}\big\langle\hat{b}_{k}\hat{b}_{l}^{\dagger}\big\rangle+e^{+i\omega_{k}t}\big\langle\hat{b}_{k}^{\dagger}\hat{b}_{l}\big\rangle\Big) \bigg[\hat{A}(\xi)\tilde{\hat{\rho}}(t)\hat{A}^{\dagger}(\xi)-\frac{1}{2}\hat{A}^{\dagger}(\xi)\hat{A}(\xi)\tilde{\hat{\rho}}(t)-\frac{1}{2}\tilde{\hat{\rho}}(t)\hat{A}^{\dagger}(\xi)\hat{A}(\xi)\bigg]. \label{eq:306}
\end{align}
Transforming the summations over the indices $k$ and $l$ into integrals, the master equation obtains the form
\begin{align}
\partial_{t}\tilde{\hat{\rho}}(t)=&\,\sum_{\xi}\int_{-\infty}^{\infty}\frac{\text{d}\omega_{1}}{2\pi}\,\rho(\omega_{1})\int_{-\infty}^{\infty}\frac{\text{d}\omega_{2}}{2\pi}\,\rho(\omega_{2})\int_{-\infty}^{\infty}\text{d}t\,e^{i\xi t} \textsl{g}(\omega_{1})\textsl{g}(\omega_{2})\Big[e^{-i\omega_{1}t}\big\langle\hat{b}(\omega_{1})\hat{b}^{\dagger}(\omega_{2})\big\rangle+e^{+i\omega_{1}t}\big\langle\hat{b}^{\dagger}(\omega_{1})\hat{b}(\omega_{2})\big\rangle\Big] \nonumber \\
&\,\times\bigg[\hat{A}(\xi)\tilde{\hat{\rho}}(t)\hat{A}^{\dagger}(\xi)-\frac{1}{2}\hat{A}^{\dagger}(\xi)\hat{A}(\xi)\tilde{\hat{\rho}}(t)-\frac{1}{2}\tilde{\hat{\rho}}(t)\hat{A}^{\dagger}(\xi)\hat{A}(\xi)\bigg], \label{eq:307}
\end{align}
where $\rho(\omega_{j})$ is the density of states. By defining $\big\langle\hat{b}(\omega_{1})\hat{b}^{\dagger}(\omega_{2})\big\rangle=2\pi\delta(\omega_{1}-\omega_{2})\big[\bar{n}(\omega_{1})+1\big]$ and $\big\langle\hat{b}^{\dagger}(\omega_{1})\hat{b}(\omega_{2})\big\rangle=2\pi\delta(\omega_{1}-\omega_{2})\bar{n}(\omega_{1})$, we finally obtain
\begingroup
\allowdisplaybreaks
\begin{align}
\partial_{t}\tilde{\hat{\rho}}(t)=&\,\sum_{\xi}\int_{-\infty}^{\infty}\frac{\text{d}\omega_{1}}{2\pi}\,\rho(\omega_{1})\int_{-\infty}^{\infty}\frac{\text{d}\omega_{2}}{2\pi}\,\rho(\omega_{2})\int_{-\infty}^{\infty}\text{d}t\,e^{i\xi t}\textsl{g}(\omega_{1})\textsl{g}(\omega_{2}) \times2\pi\delta(\omega_{1}-\omega_{2})\Big\{ e^{-i\omega_{1}t}\big[\bar{n}(\omega_{1})+1\big]+e^{+i\omega_{1}t}\bar{n}(\omega_{1})\Big\} \nonumber \\
&\,\times\bigg[\hat{A}(\xi)\tilde{\hat{\rho}}(t)\hat{A}^{\dagger}(\xi)-\frac{1}{2}\hat{A}^{\dagger}(\xi)\hat{A}(\xi)\tilde{\hat{\rho}}(t)-\frac{1}{2}\tilde{\hat{\rho}}(t)\hat{A}^{\dagger}(\xi)\hat{A}(\xi)\bigg] \nonumber \\
=&\,\sum_{\xi}\int_{-\infty}^{\infty}\frac{\text{d}\omega_{1}}{2\pi}\,\rho^{2}(\omega_{1})\int_{-\infty}^{\infty}\text{d}t\,e^{i\xi t}\textsl{g}^{2}(\omega_{1})\Big\{ e^{-i\omega_{1}t}\big[\bar{n}(\omega_{1})+1\big]+e^{+i\omega_{1}t}\bar{n}(\omega_{1})\Big\} \nonumber \\
&\,\times\bigg[\hat{A}(\xi)\tilde{\hat{\rho}}(t)\hat{A}^{\dagger}(\xi)-\frac{1}{2}\hat{A}^{\dagger}(\xi)\hat{A}(\xi)\tilde{\hat{\rho}}(t)-\frac{1}{2}\tilde{\hat{\rho}}(t)\hat{A}^{\dagger}(\xi)\hat{A}(\xi)\bigg] \nonumber \\
=&\,\sum_{\xi}\int_{-\infty}^{\infty}\text{d}\omega_{1}\,\rho^{2}(\omega_{1})\textsl{g}^{2}(\omega_{1})\Big\{\delta(\xi-\omega_{1})\big[\bar{n}(\omega_{1})+1\big]+\delta(\xi+\omega_{1})\bar{n}(\omega_{1})\Big\} \nonumber \\
&\,\times\bigg[\hat{A}(\xi)\tilde{\hat{\rho}}(t)\hat{A}^{\dagger}(\xi)-\frac{1}{2}\hat{A}^{\dagger}(\xi)\hat{A}(\xi)\tilde{\hat{\rho}}(t)-\frac{1}{2}\tilde{\hat{\rho}}(t)\hat{A}^{\dagger}(\xi)\hat{A}(\xi)\bigg] \nonumber \\
=&\,\sum_{\xi}\rho^{2}(\xi)\textsl{g}^{2}(\xi)\big[\bar{n}(\xi)+1\big]\bigg[\hat{A}(\xi)\tilde{\hat{\rho}}(t)\hat{A}^{\dagger}(\xi)-\frac{1}{2}\hat{A}^{\dagger}(\xi)\hat{A}(\xi)\tilde{\hat{\rho}}(t)-\frac{1}{2}\tilde{\hat{\rho}}(t)\hat{A}^{\dagger}(\xi)\hat{A}(\xi)\bigg] \nonumber \\
&\,+\sum_{\xi}\rho^{2}(-\xi)\textsl{g}^{2}(-\xi)\bar{n}(-\xi)\bigg[\hat{A}(\xi)\tilde{\hat{\rho}}(t)\hat{A}^{\dagger}(\xi)-\frac{1}{2}\hat{A}^{\dagger}(\xi)\hat{A}(\xi)\tilde{\hat{\rho}}(t)-\frac{1}{2}\tilde{\hat{\rho}}(t)\hat{A}^{\dagger}(\xi)\hat{A}(\xi)\bigg] \nonumber \\
=&\,\sum_{\xi>0}\rho^{2}(\xi)\textsl{g}^{2}(\xi)\big[\bar{n}(\xi)+1\big]\bigg[\hat{A}(\xi)\tilde{\hat{\rho}}(t)\hat{A}^{\dagger}(\xi)-\frac{1}{2}\hat{A}^{\dagger}(\xi)\hat{A}(\xi)\tilde{\hat{\rho}}(t)-\frac{1}{2}\tilde{\hat{\rho}}(t)\hat{A}^{\dagger}(\xi)\hat{A}(\xi)\bigg] \nonumber \\
&\,+\sum_{\xi<0}\rho^{2}(-\xi)\textsl{g}^{2}(-\xi)\bar{n}(-\xi)\bigg[\hat{A}(\xi)\tilde{\hat{\rho}}(t)\hat{A}^{\dagger}(\xi)-\frac{1}{2}\hat{A}^{\dagger}(\xi)\hat{A}(\xi)\tilde{\hat{\rho}}(t)-\frac{1}{2}\tilde{\hat{\rho}}(t)\hat{A}^{\dagger}(\xi)\hat{A}(\xi)\bigg] \nonumber \\
=&\,\sum_{n=0}^{\infty}\rho^{2}(\omega+2\eta n)\textsl{g}^{2}(\omega+2\eta n)\big[\bar{n}(\omega+2\eta n)+1\big](n+1)\bigg[|n\rangle\langle n+1|\tilde{\hat{\rho}}(t)|n+1\rangle\langle n|-\frac{1}{2}\Big\{|n+1\rangle\langle n+1|,\tilde{\hat{\rho}}(t)\Big\}\bigg] \nonumber \\
&\,+\sum_{n=0}^{\infty}\rho^{2}(\omega+2\eta n)\textsl{g}^{2}(\omega+2\eta n)\bar{n}(\omega+2\eta n)(n+1)\bigg[|n+1\rangle\langle n|\tilde{\hat{\rho}}(t)|n\rangle\langle n+1|-\frac{1}{2}\Big\{|n\rangle\langle n|,\tilde{\hat{\rho}}(t)\Big\}\bigg] \nonumber \\
\simeq&\,\frac{\kappa}{2}\sum_{n=0}^{\infty}\big(\bar{n}_{n}+1\big)(n+1)\Big[2|n\rangle\langle n+1|\tilde{\hat{\rho}}(t)|n+1\rangle\langle n|-\Big\{|n+1\rangle\langle n+1|,\tilde{\hat{\rho}}(t)\Big\}\Big] \nonumber \\
&\,+\frac{\kappa}{2}\sum_{n=0}^{\infty}\bar{n}_{n}(n+1)\Big[2|n+1\rangle\langle n|\tilde{\hat{\rho}}(t)|n\rangle\langle n+1|-\Big\{|n\rangle\langle n|,\tilde{\hat{\rho}}(t)\Big\}\Big], \label{eq:308}
\end{align}
\endgroup
where we have used the fact that the density of states is zero for negative frequencies, approximated $\rho(\omega+2\eta n)\approx\rho(\omega)$ and $\textsl{g}(\omega+2\eta n)\approx \textsl{g}(\omega)$, and defined $\kappa :=\rho^{2}(\omega)\textsl{g}^{2}(\omega)$ and
\begin{align}
\bar{n}_{n}:=\bar{n}(\omega+2\eta n)=&\, \text{Tr}\Big[\hat{b}_n^{\dagger}\hat{b}_n e^{-\hbar(\omega+2n\eta ) \hat{b}_n^{\dagger}\hat{b}_n/k_{\text{B}}T_{\text{E}}}/Z \Big] = \frac{1}{e^{\hbar(\omega+2n\eta )/k_{\text{B}}T_{\text{E}}}-1}. \label{eq:310}
\end{align}
Finally, transforming back to the Schrödinger picture according to $e^{-i\hat{H}t/\hbar}\tilde{\hat{\rho}}(t)e^{+i\hat{H}t/\hbar}=\hat{\rho}(t)$, we have
\begin{align}
\partial_{t}\hat{\rho}(t)=&\,\frac{i}{\hbar}\big[\hat{\rho}(t),\hat{H}\big]+\frac{\kappa}{2}\sum_{n=0}^{\infty}\big(\bar{n}_{n}+1\big)(n+1)\Big[2|n\rangle\langle n+1|\hat{\rho}(t)|n+1\rangle\langle n|-\Big\{|n+1\rangle\langle n+1|,\hat{\rho}(t)\Big\}\Big] \nonumber \\
&\,+\frac{\kappa}{2}\sum_{n=0}^{\infty}\bar{n}_{n}(n+1)\Big[2|n+1\rangle\langle n|\hat{\rho}(t)|n\rangle\langle n+1|-\Big\{|n\rangle\langle n|,\hat{\rho}(t)\Big\}\Big], \label{eq:309}
\end{align}
where $T_{\text{E}}$ is the temperature of the environment.

\textbf{Dynamic equation for the generating function for the dissipator}---Here, we derive the dynamic equation for the dissipator of the anharmonic oscillator derived above. For simplicity, we assume zero-temperature bath such that $\bar{n}_{n}=0$ for all $n\in\mathbb{N}_{0}$. We begin by writing down the adjoint of the dissipator part of Eq.~(\ref{eq:309}),
\begin{align}
\partial_{t}\hat{A}(t)=&\,\frac{\kappa}{2}\sum_{n=0}^{\infty}(n+1)\Big[2|n+1\rangle\langle n|\hat{A}(t)|n\rangle\langle n+1|-|n+1\rangle\langle n+1|\hat{A}(t)-\hat{A}(t)|n+1\rangle\langle n+1|\Big] \nonumber \\
=&\,\frac{\kappa}{2}\Bigg[2\sum_{n=0}^{\infty}(n+1)|n+1\rangle\langle n|\hat{A}(t)|n\rangle\langle n+1|-\hat{a}^{\dagger}\hat{a}\hat{A}(t)-\hat{A}(t)\hat{a}^{\dagger}\hat{a}\Bigg]. \label{eq:311}
\end{align}
By expanding the operator under interest in the normal order this equation obtains the form
\begin{align}
\partial_{t}\sum_{k,l=0}^{\infty}&f_{k,l}(t)\big(\hat{a}^{\dagger}\big)^{k}\hat{a}^{l} \nonumber \\
=&\,\frac{\kappa}{2}\Bigg[2\sum_{n=0}^{\infty}(n+1)|n+1\rangle\langle n|\sum_{k,l=0}^{\infty}f_{k,l}(t)\big(\hat{a}^{\dagger}\big)^{k}\hat{a}^{l}|n\rangle\langle n+1| -\hat{a}^{\dagger}\hat{a}\sum_{k,l=0}^{\infty}f_{k,l}(t)\big(\hat{a}^{\dagger}\big)^{k}\hat{a}^{l}-\sum_{k,l=0}^{\infty}f_{k,l}(t)\big(\hat{a}^{\dagger}\big)^{k}\hat{a}^{l}\hat{a}^{\dagger}\hat{a}\Bigg] \nonumber \\
=&\,\frac{\kappa}{2}\Bigg\{2\sum_{n=0}^{\infty}(n+1)\sum_{k=0}^{\infty}f_{k,k}(t)\frac{n!}{(n-k)!}|n+1\rangle\langle n+1| \nonumber \\
&\,-\hat{a}^{\dagger}\sum_{k,l=0}^{\infty}f_{k,l}(t)\Big[\big(\hat{a}^{\dagger}\big)^{k}\hat{a}+k\big(\hat{a}^{\dagger}\big)^{k-1}\Big]\hat{a}^{l}-\sum_{k,l=0}^{\infty}f_{k,l}(t)\big(\hat{a}^{\dagger}\big)^{k}\big(\hat{a}^{\dagger}\hat{a}^{l}+l\hat{a}^{l-1}\big)\hat{a}\Bigg\} \nonumber \\
=&\,\frac{\kappa}{2}\Bigg[2\sum_{n=0}^{\infty}\sum_{k=0}^{\infty}f_{k,k}(t)\frac{(n+1)!}{(n-k)!}|n+1\rangle\langle n+1| -2\sum_{k,l=0}^{\infty}f_{k,l}(t)\big(\hat{a}^{\dagger}\big)^{k+1}\hat{a}^{l+1}-\sum_{k,l=0}^{\infty}f_{k,l}(t)(k+l)\big(\hat{a}^{\dagger}\big)^{k}\hat{a}^{l}\Bigg] \nonumber \\
=&\,\frac{\kappa}{2}\Bigg[2\sum_{n=0}^{\infty}\sum_{k=0}^{\infty}f_{k,k}(t)\big(\hat{a}^{\dagger}\big)^{k+1}\hat{a}^{k+1}|n+1\rangle\langle n+1| -2\sum_{k,l=0}^{\infty}f_{k,l}(t)\big(\hat{a}^{\dagger}\big)^{k+1}\hat{a}^{l+1}-\sum_{k,l=0}^{\infty}f_{k,l}(t)(k+l)\big(\hat{a}^{\dagger}\big)^{k}\hat{a}^{l}\Bigg] \nonumber \\
=&\,\frac{\kappa}{2}\Bigg[2\sum_{k=0}^{\infty}f_{k,k}(t)\big(\hat{a}^{\dagger}\big)^{k+1}\hat{a}^{k+1}\sum_{n=0}^{\infty}|n\rangle\langle n| -2\sum_{k,l=0}^{\infty}f_{k,l}(t)\big(\hat{a}^{\dagger}\big)^{k+1}\hat{a}^{l+1}-\sum_{k,l=0}^{\infty}f_{k,l}(t)(k+l)\big(\hat{a}^{\dagger}\big)^{k}\hat{a}^{l}\Bigg] \nonumber \\
=&\,\frac{\kappa}{2}\Bigg[2\sum_{k,l=0}^{\infty}f_{k-1,k-1}(t)\delta_{k,l}\big(\hat{a}^{\dagger}\big)^{k}\hat{a}^{l} -2\sum_{k,l=0}^{\infty}f_{k-1,l-1}(t)\big(\hat{a}^{\dagger}\big)^{k}\hat{a}^{l}-\sum_{k,l=0}^{\infty}f_{k,l}(t)(k+l)\big(\hat{a}^{\dagger}\big)^{k}\hat{a}^{l}\Bigg]. \label{eq:312}
\end{align}
The completeness of the normal ordered monomials yields
\begin{align}
\dot{f}_{k,l}(t)=&\,\frac{\kappa}{2}\big[2f_{k-1,k-1}(t)\delta_{k,l}-2f_{k-1,l-1}(t)-f_{k,l}(t)(k+l)\big]. \label{eq:313}
\end{align}
Multiplying both sides by $x^{k}y^{l}$, performing summations over $k$ and $l$, and defining $F(x,y;t)=\sum_{k,l}f_{k,l}(t)x^{k}y^{l}$ as before gives
\begin{align}
\sum_{k,l=0}^{\infty}\dot{f}_{k,l}(t)x^{k}y^{l}=&\,\frac{\kappa}{2}\Bigg[2\sum_{k,l=0}^{\infty}f_{k-1,k-1}(t)\delta_{k,l}x^{k}y^{l}-2\sum_{k,l=0}^{\infty}f_{k-1,l-1}(t)x^{k}y^{l}-\sum_{k,l=0}^{\infty}f_{k,l}(t)(k+l)x^{k}y^{l}\Bigg] \nonumber \\
=&\,\frac{\kappa}{2}\Bigg[2xy\sum_{k,l=0}^{\infty}f_{k,k}(t)\delta_{k,l}x^{k}y^{l}-2xy\sum_{k,l=0}^{\infty}f_{k,l}(t)x^{k}y^{l}-\sum_{k,l=0}^{\infty}f_{k,l}(t)(k+l)x^{k}y^{l}\Bigg] \nonumber \\
\Rightarrow\dot{F}(x,y;t)=&\,\frac{\kappa}{2}\big(2P-2xy-x\partial_{x}-y\partial_{y}\big) F(x,y;t), \label{eq:314}
\end{align}
where $P F(x,y;t)=xy\sum_{k=0}^{\infty} f_{k,k}(t)(xy)^k$. Defining the transformed generating function as in the main text, $G(x,y;t)=e^{xy}F(x,y;t)$, finally gives
\begin{align}
\dot{G}(x,y;t)=&\,\frac{\kappa}{2}\big(2P- x\partial_{x}-y\partial_{y}\big) G(x,y;t), \label{eq:315}
\end{align}
where $P G(x,y;t)=xy\sum_{k,l=0}^{\infty}\delta_{k,l} g_{k,l}(t)x^k y^l$.

\section{Entropy production rate of the anharmonic oscillator}
\label{sec:ent}

\textbf{Conventional master equation}---Consider the conventionally-used master equation for an anharmonic oscillator coupled to a Markovian heat bath~\cite{Chaturvedi_1991}
\begin{align}
\dot{\hat{\rho}}(t) =&\,-i\omega[\hat{a}^{\dagger}\hat{a},\hat{\rho}(t)] - i \eta[(\hat{a}^{\dagger})^2\hat{a}^2,\hat{\rho}(t)] \nonumber \\
&\,+ \frac{\kappa}{2}(\bar{n}+1)\big[2\hat{a}\hat{\rho}(t)\hat{a}^{\dagger} - \hat{a}^{\dagger}\hat{a}\hat{\rho}(t) - \hat{\rho}(t)\hat{a}^{\dagger}\hat{a}\big] + \frac{\kappa}{2}\bar{n}\big[2\hat{a}^{\dagger}\hat{\rho}(t)\hat{a} - \hat{a}\hat{a}^{\dagger}\hat{\rho}(t) - \hat{\rho}(t)\hat{a}\hat{a}^{\dagger}\big], \label{eq:201}
\end{align}
where $\bar{n}=(e^{\hbar \omega/k_{\text{B}}T_{\text{E}}}-1)^{-1}$. Recall the definition of the entropy production rate~\cite{Breuer_2002},
\begin{align}
\Pi(t) =&\,\dot{S}(t) - \frac{1}{T_{\text{E}}} \dot{E}(t) \label{eq:202} \\
=&\,-k_{\text{B}}\partial_t \text{Tr}\big[ \hat{\rho}(t) \ln \hat{\rho}(t)\big] - \frac{1}{T_{\text{E}}} \partial_t \text{Tr}\big[ \hat{H}\hat{\rho}(t) \big] \nonumber \\
=&\,-k_{\text{B}}\text{Tr}\big[ \dot{\hat{\rho}}(t) \ln \hat{\rho}(t)\big] - k_{\text{B}}\text{Tr}\big[ \dot{\hat{\rho}}(t)\big] - \frac{1}{T_{\text{E}}} \text{Tr}\big[ \hat{H}\dot{\hat{\rho}}(t) \big]. \label{eq:203}
\end{align}
Consider a diagonal Gaussian initial state, $\hat{\rho}(0)=\mathcal{N} \sum_{n=0}^{\infty} e^{-cn}|n\rangle \langle n|$, where $\mathcal{N}=1-e^{-c}$ is a normalization constant and $c\in \mathbb{R}_+$ is a constant that characterizes the state. Direct insertion of the state and Eq.~(\ref{eq:201}) into Eq.~(\ref{eq:203}) yields for the scaled entropy production rate at $t=0$ 
\begin{align}
T_{\text{E}}\Pi(0)/\hbar\kappa=&\, (N_{0}-\bar{n} )\big[\omega\big(1-T_{\text{E}}/T_{0}\big)+4\eta N_{0}\big], \label{eq:204}
\end{align}
where we have defined $N_{0}=(e^c-1)^{-1}$ and $T_0=\hbar \omega/k_{\text{B}}c$.

Since $T_{\text{E}}, \kappa > 0$, Eq.~(\ref{eq:204}) implies that the initial entropy production is negative iff
\begin{subequations}
\begin{align}
N_{0}-\bar{n}>&\,0, \label{eq:205a} \\
\omega\big(1-T_{\text{E}}/T_{0}\big)+4\eta N_{0}<&\,0, \label{eq:205b}
\end{align}
\end{subequations}
or
\begin{subequations}
\begin{align}
N_{0}-\bar{n}<&\,0, \label{eq:206a} \\
\omega\big(1-T_{\text{E}}/T_{0}\big)+4\eta N_{0}>&\,0. \label{eq:206b}
\end{align}
\end{subequations}
First, if $N_{0}-\bar{n}>0$, then Eqs.~(\ref{eq:205a}) and~(\ref{eq:205b}) are satisfied iff
\begin{subequations}
\begin{align}
T_{\text{E}}<T_0<&\, \frac{T_{\text{E}}}{4N_0\eta/\omega +1}, \label{eq:207a} \\
\omega + 4 \eta N_0>&\, 0, \label{eq:207b}
\end{align}
\end{subequations}
or
\begin{subequations}
\begin{align}
T_{\text{E}}<&\, T_0, \label{eq:208a} \\
\omega + 4 \eta N_0\leq&\, 0. \label{eq:208b}
\end{align}
\end{subequations}
This implies that for all $\eta<0$ exists such $T_0$, that is, initial state, that the initial entropy production rate is negative.

Second, if $N_{0}-\bar{n}<0$, then Eqs.~(\ref{eq:206a}) and~(\ref{eq:206b}) are satisfied iff
\begin{subequations}
\begin{align}
\frac{T_{\text{E}}}{4N_0\eta/\omega +1}<T_0<&\, T_{\text{E}}, \label{eq:209a} \\
\omega + 4 \eta N_0>&\, 0. \label{eq:209b}
\end{align}
\end{subequations}
Note that Eq.~(\ref{eq:209b}) is satisfied for all $\eta>0$ since $\omega, N_0>0$. Equation~(\ref{eq:209a}) implies that for all $\eta>0$ exists such an initial state that the initial entropy production rate is negative.

Collecting the above results together, we conclude that for any finite anharmonicity, $\eta\neq 0$, there exists an initial state that leads to negative entropy production rate in an anharmonic oscillator governed by the conventionally-used master equation, Eq.~(\ref{eq:201}).

\textbf{Master equation derived from the first principles}---Consider a diagonal Gaussian initial state as in the previous section. Moreover, assume low temperature for the environment and for the initial state, such that $\hbar \omega /k_{\text{B}}T_{\text{E}} \gg 1$, $\hbar \eta /k_{\text{B}}T_{\text{E}} \gg 1$, and $\hbar \omega /k_{\text{B}}T_0 \gg 1$. Again, direct insertion of the state and Eq.~(\ref{eq:309}) into Eq.~(\ref{eq:203}) yields for the scaled entropy production rate at $t=0$ 
\begin{align}
T_{0}T_{\text{E}}\Pi(0)/(\hbar\kappa\omega)\approx&\,\big(T_{\text{E}}-T_{0}\big)\big(e^{-\hbar\omega/k_{\text{B}}T_{\text{E}}}-e^{-\hbar\omega/k_{\text{B}}T_{0}}\big). \label{eq:210}
\end{align}
Evidently, both of the factors on the right-hand side of the equation have the same sign. Consequently, $\Pi(0)\geq 0$ for all choices of the parameters in this limit. We conclude that the master equation derived from the first principles corrects the contradiction between the conventionally-used Markovian master equation and the well-known property of Markovian master equations that the entropy production rate is non-negative~\cite{Breuer_2002}, at least for the setup considered here.

\begin{figure}[b!]
	\centering
	\includegraphics[width = 0mm]{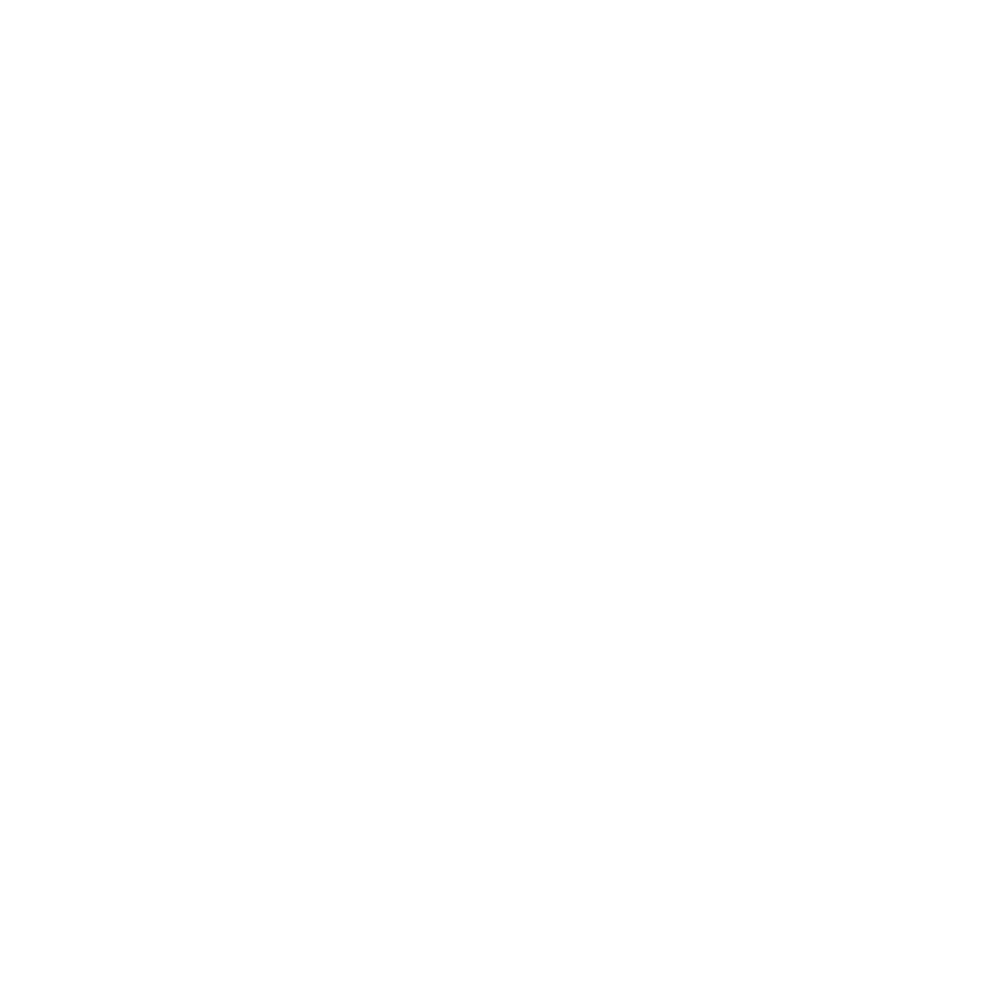} 
\end{figure}

\bibliographystyle{apsrev4-1}
\bibliography{supplemental}